# Non-monotonic temperature dependence of electron viscosity and crossover to high-temperature universal viscous fluid in monolayer and bilayer graphene

Indra Yudhistira,[1,2] Ramal Afrose,[1,2] and Shaffique Adam[1,2,3,4,5]

[1] *Centre for Advanced 2D Materials, National University of Singapore, 6 Science Drive 2, 117546, Singapore*
[2] *Department of Physics, National University of Singapore, 2 Science Drive 3, 117551, Singapore*
[3] *Department of Materials Science and Engineering, National University of Singapore, 9 Engineering Drive 1, Singapore 117575*
[4] *Yale-NUS College, 16 College Avenue West, 138527, Singapore*
[5] *Department of Physics, Washington University in St. Louis, St. Louis, Missouri 63130, United States*

Electrons in quantum matter behave like a fluid when the quantum-mechanical carrier-carrier scattering dominates over other relaxation mechanisms. By combining a microscopic treatment of electron-electron interactions within the random phase approximation with a phenomenological Navier-Stokes like equation, we predict that in the limit of high temperature and strong Coulomb interactions, both monolayer and bilayer graphene exhibit a universal behavior in dynamic viscosity. We find that the dynamic viscosity to entropy density ratio for bilayer graphene is closer to the holographic bound suggesting that such a bound might be observable in a condensed matter system. We discuss how this could be observed experimentally using a magnetoconductance measurements in a Corbino geometry for a realistic range of temperature and carrier density.

## I. INTRODUCTION

Most metals typically exhibit diffusive or ballistic transport behavior, due to strong electron-impurity scattering, electron-phonon scattering or scattering with boundaries. However, in instances where carrier-carrier scattering surpasses all other forms of scattering, a hydrodynamic regime emerges, characterized by electronic transport dominated by viscous effects [1]. This regime holds particular significance for two dimensional materials, in particular graphene and bilayer graphene, for which the electron-electron interactions are intrinsically strong. Moreover, recent advancements in experimental techniques have enabled the creation of ultra-clean sample devices with minimal disorder. The experimental observation of electron hydrodynamics has been sparse for many years [2, 3]. However, recently there has been a rapid experimental progress. Reports of electronic hydrodynamic behavior have surfaced in monolayer graphene [4–13], bilayer graphene [4, 14, 15], PdCoO [16], GaAs/AlGaAs [17–22], $WP_2$[23] and $WTe_2$ [24, 25], prompting the development of new theories to elucidate these phenomena [26–36]. In particular, in a graphene Corbino devices at charge neutrality, a strong magnetoresistance at a relatively weak field [33] and anomalous thermal transport [35] has been predicted and the hydrodynamic flow is dominated by energy current [36].

In the hydrodynamic regime, the electron fluid is described by the Navier-Stokes equation as an effective long-wavelength theory. An important parameter in this description is the dynamic viscosity $\eta$ that describes the momentum transfer through the fluid with an applied sheer stress ($\eta$ is formally defined as the ratio of the shear stress to shear rate). The dynamic viscosity is of particular interest to us because it can be conveniently measured in magnetotransport experiments serving as the bridge between theory and experimental observations (see e. g. [30, 31, 37] for details). The dynamic viscosity also connects hydrodynamic electrons in solid state materials to more exotic systems. Motivated by the nearly perfect fluid behavior seen in relativistic heavy ion collider experiments, Kovtun et. al. [26] proposed a lower bound for the ratio of dynamic shear viscosity to entropy density for a large class of strongly interacting quantum field theories using "holographic correspondence". This inspired Muller et. al. [28] to show that clean undoped monolayer graphene is a nearly perfect fluid owing to the low viscosity to entropy density ratio whose computed value is within a factor of $4\pi$ from the lower bound proposed for a perfect fluid [26]. In a separate theoretical work, Son showed that despite the modification of graphene quasiparticle properties by the long-range Coulomb interaction [38], in the limit of infinitely strong Coulomb interaction, the 2D Dirac Fermion system hosts stable quasiparticles [27]. This is because quasiparticle decay into two or more other quasiparticles is forbidden by energy and momentum conservation. The implication is that the hydrodynamic fluid in monolayer graphene is stable to strong Coulomb interactions. However, a similar consideration for bilayer graphene remains unclear [39, 40].

In this work, we demonstrate that the dynamic shear viscosity for both monolayer and bilayer graphene exhibits a non-monotonic temperature dependence. The low-temperature degenerate regime gives a density-dependent viscosity that decreases with increasing temperature. At high temperature, there is a crossover to a universal limit that is independent of both carrier-density and interaction-strength and characterized by a universal dynamic viscosity to entropy density ratio. While the stability of monolayer graphene to strong Coulomb interactions was already established [27], we find that for bilayer graphene, in the absence of a band-gap, this phase is stable with increasing Coulomb interactions. However, in the presence of a band-gap, for sufficiently strong Coulomb interactions the universal fluid is unstable to exciton formation. Nonetheless, we demonstrate a wide range of experimentally accessible carrier densities and temperature over which the universal behavior should be observable. In order to establish a connection with experiment, we suggest a protocol to experimentally measure the dynamical viscosity through magnetoconductance measurements in a Corbino geometry.

The dynamic viscosity for an electronic material can be calculated using different approaches. We can very broadly categorize these in two families: (i) linear response approaches and, and (ii) the Boltzmann transport method. For the linear response method, the dynamic viscosity is determined from the low frequency stress-stress response function in the spirit of the Kubo formula [41]. It is given by $\eta = -\lim_{\omega\to 0}(1/(4\omega))\sum_{i,j}\mathcal{I}m[\chi_{ij,ij}(0,\omega) - (1/2)\chi_{ii,jj}(0,\omega)]$, where $\chi_{ij,kl}(\boldsymbol{q},\omega)$ is the stress-stress response function [42]. The linear response approach will necessarily involve approximations in terms of diagrammatic representation of the stress-stress response function as well as retaining the singular part of the Green's function that corresponds to what is known as the quasiparticle approximation in the spirit of the theory of normal Fermi liquid. For monolayer graphene, this approach was used by Ref. [42].

For the Boltzmann transport method, the central quantity to be solved is the non-equilibrium electronic distribution function. In this way, the dynamic viscosity is obtained in terms of electronic distribution function from $\delta\Pi^{\alpha\beta} = \sum_{\lambda=e,h} g\int d^2k/(2\pi)^2\hbar v^{\alpha}_{\boldsymbol{k},e(h)}k^{\beta}\delta f_{\mathbf{k},\lambda} = \eta\nabla^2\mathbf{u} + (\eta/2)\nabla(\nabla\cdot\boldsymbol{u})$, where $\delta\Pi_{ij}$ and $\delta f_{\mathbf{k},\lambda}$ is the non-equilibrium part of the stress-energy tensor and electronic distribution function respectively. The non-equilibrium electronic distribution function is typically obtained through a variational solution where one introduces basis functions that capture the symmetry of the problem. This approach has been done for the 2D electron gas using a single basis function in Refs. [43, 44], and for monolayer graphene Ref. [45] used three basis functions that they dubbed as the "there-mode" approximations since it showed the existence of three macroscopic currents carrying electric charge, energy, and quasiparticle imbalance.

For the 2D electron gas, both approaches give $\eta = m^\star n(1/4)v_F^2\tau_\eta$, where $m^\star$ is the effective mass, $n$ is the quasiparticle density, $v_F$ is the quasiparticle group velocity which is $T$ independent, and $\tau_\eta$ is the timescale for viscous relaxation. Similarly, for monolayer graphene, Ref. [42] found that at $T \ll T_F$, the dynamic viscosity can be written as $\eta = (1/4)nE_F\tau_\eta$, where $E_F$ is the Fermi energy. One observes that for monolayer graphene, $E_F/v_F^2$ plays a similar role to $m^\star$ in 2D parabolic system. Based on this insight, we postulate that the relation of $\eta = \rho_m(1/4)v_F^2\tau_\eta$ holds even for materials without a linear or parabolic band provided we calculate the effective mass density $\rho_m$ appropriate for the corresponding energy dispersion. Moreover, in the 2D electron gas, it was also known that the viscous transport time goes as $\tau_\eta/\hbar \sim [E_F/(k_BT)^2][\ln(T_F/T)]^2$ [43] while the quasiparticle lifetime goes as $\tau_{\rm qp}/\hbar \sim [E_F/(k_BT)^2]/\ln(T_F/T)$ [46], both having similar leading order behavior in $T/T_F$. Based on this insight, we also postulate that the quasiparticle lifetime will continue to be a good approximation of viscous transport time for non-parabolic bands.

We use this approach because it gives us insight to the origin of the temperature and carrier density dependence of dynamic viscosity and the interplay between effective mass density and kinematic viscosity (hence viscous transport time) in the same way as the 2D parabolic system. In addition, it frees us from the heavy numerics of solving self-consistent Bethe-Salpeter equation and inaccuracy coming from the choice of limited basis function used to expand the non-equilibrium distribution function in solving Boltzmann equation.

The rest of this paper is organized as follows: In Sec. II, we formulate the microscopic model underlying our theory. After discussing the framework, we calculate the dependence of the dynamic viscosity on temperature and Coulomb interactions in Sec. III. We then discuss the stability of the universal hydrodynamic fluid in Sec. IV studying the possibility of exciton formation with strong Coulomb interactions. In Sec. V we show that the dynamical viscosity can be measured experimentally using magnetoconductance measurements in a Corbino geometry. Finally, we conclude in Sec. VI.

## II. MICROSCOPIC MODEL

The starting point for the microscopic calculation of dynamic viscosity is the electron-electron lifetime $\tau_{ee}$ obtained from energy dependent quasiparticle lifetime $\tau(\varepsilon_{\boldsymbol{k},\lambda}) = \hbar/2{\rm Im}[\Sigma_\lambda(\boldsymbol{k},\varepsilon_{\boldsymbol{k},\lambda})]$, where $\Sigma_\lambda(\boldsymbol{k},\omega)$ is the corresponding self-energy calculated under random phase approximation (see e.g. Ref. [47] for details). We reproduced here imaginary parts of the self-energy for completeness

$$\Im m[\Sigma_\lambda(\boldsymbol{k},\omega)] = -\int\frac{d^2\boldsymbol{q}}{(2\pi)^2}\sum_{\lambda'}\Im m\left[\frac{V(q)}{\epsilon(q,\omega-\varepsilon_{\boldsymbol{k}-\boldsymbol{q},\lambda'}+\mu,T)}\right]\mathcal{F}_{\lambda\lambda'}(\theta_{\boldsymbol{k},\boldsymbol{k}-\boldsymbol{q}})\left[n_{\rm B}(\omega-\varepsilon_{\boldsymbol{k}-\boldsymbol{q},\lambda'}+\mu)+n_{\rm F}(-\varepsilon_{\boldsymbol{k}-\boldsymbol{q},\lambda'}+\mu)\right], \tag{1}$$

where $V(q) = 2\pi e^2/(\kappa q)$ is the Coulomb potential, $n_{\mathrm{B/F}}(x) \equiv 1/[\exp(\beta x) \mp 1]$ the Bose (Fermi) distribution function, $\lambda, \lambda'$ are band indices ($\pm$ denote the conduction and valence bands), $\mathcal{F}_{\lambda\lambda'}(\varphi)$ is the chirality factor, and $\varphi$ is the scattering angle. $\epsilon(q,\omega,T) \equiv 1 - V(q)\Pi(q,\omega,T)$ refers to the dynamical RPA dielectric function (see Ref. [48] and [49] for dynamical RPA polarizability of monolayer and bilayer graphene respectively).

We thermally averaged this quantity by the procedure laid out in [50], and reproduced here for completeness

$$\langle \tau^{-1} \rangle = \frac{\int_{-\infty}^{\infty} d\varepsilon\ |\varepsilon|\ \frac{\partial f^{(0)}(\varepsilon-\mu)}{\partial \varepsilon}\ \tau^{-1}(\varepsilon)}{\int_{-\infty}^{\infty} d\varepsilon\ |\varepsilon|\ \frac{\partial f^{(0)}(\varepsilon-\mu)}{\partial \varepsilon}}, \tag{2}$$

This procedure then allows us to obtain the kinematic viscosity [51], $\nu = (1/4)v^2(n,T)\tau_{ee,2} \approx v^2(n,T)\tau_{ee}$, where $v$ is the effective velocity, defined here as $(1/\hbar)\partial\varepsilon/\partial k$ with the wavevector $k$ replaced by $\sqrt{\pi(n_e+n_h)}$ thus taking thermal fluctuations into account, and $\tau_{ee}$ is a shorthand of $\langle\tau^{-1}\rangle^{-1}$. Here we have used the relation between viscosity transport time and lifetime $\tau_{ee,2} \approx 4\tau_{ee}$, which we explained in more detail in the Appendix. A.

In this work, we use Dirac model having energy dispersion $\varepsilon_{\boldsymbol{k},\pm} = \pm\hbar v_F k$ for monolayer graphene. The Dirac dispersion breaks down at high energy which we can estimate as 12000 K. Beyond that energy, one would need to use a tight binding model on the honeycomb lattice. Similarly, for Bernal stacked bilayer graphene, we use a double parabolic energy dispersion of $\varepsilon_{\boldsymbol{k},\pm} = \pm\hbar^2k^2/2m^\star$ for AB-stacked bilayer graphene, where $m^\star$ is the effective mass. This model breaks down at both low energy where in the presence of an electric field, one gets a Mexican hat spectrum, and also at high energy where the band starts to become linear. We can estimate that for energies $2\times10^{-3}$ eV $\lesssim \varepsilon \lesssim 0.1$ eV bilayer graphene has a quadratic spectrum [52, 53]. We also note despite being in the hydrodynamic regime $\tau_e e \lesssim \tau_i mp$, there is still a finite number of charged impurities in the substrate that induce a residual density $n^*$ in bilayer graphene that corresponds to a typical Fermi energy $\varepsilon_F \gtrsim 0.01$ eV which is typically much larger than the lower energy scale for using the quadratic Hamiltonian for the range of experimental gate voltages. See e. g. Ref. [15] where this point was discussed even in the presence of strong electric fields. The gate voltage of $V_g \lesssim 50\ V$ induces a maximum carrier density with Fermi energy $\varepsilon_F \lesssim 0.1$ eV which is comparable to the limit where the low-energy effective quadratic Hamiltonian begins to break down [54]. Therefore, for realistic samples, the condition $2\times10^{-3}$ eV $\lesssim \varepsilon_F \lesssim 0.1$ eV is mostly satisfied and the quadratic energy dispersion is an excellent approximation.

For monolayer graphene the effective velocity is given by the Fermi velocity $v_F$, which is independent of both density and temperature. However, for bilayer graphene we need to take the temperature dependence that comes from electron-hole plasma into account, i. e. $v(n,T) = (\hbar/m^\star)\sqrt{\pi(n_e+n_h)}$. This temperature dependent carrier density of the bilayer can be calculated as

$$\begin{aligned} n_e + n_h &= \sum_{\lambda=e,h} g \int \frac{d^2k}{(2\pi)^2} f_{\boldsymbol{k},\lambda} \\ &= \frac{2m^\star k_B T}{\pi\hbar^2} \sum_{\lambda=\pm1} \ln\left[1 + e^{\lambda\varepsilon_F/(k_BT)}\right] \qquad ;\text{bilayer}, \end{aligned} \tag{3}$$

where we have used Fermi function for electron(holes) given by $f^{(0)}_{\boldsymbol{k},e(h)} = 1/\{\exp[\beta(\varepsilon_{\boldsymbol{k},e(h)} \mp \mu)] + 1]\}$ and $g$ accounts for valley and spin degeneracy. At high temperature or low density ($T \gg T_F$), the effective Fermi velocity for bilayer graphene is given by $v_F(T \gg T_F) \approx \sqrt{(4\ln 2)k_BT/m^\star}$, while in the opposite limit, it is given by $v_F(T \ll T_F) \approx \sqrt{2|\varepsilon_F|/m^\star}$. This latter result is similar to the 2D electron gas [55] and classical fluids [56].

The role of kinematic viscosity in the Navier-Stokes equation for an electron fluid is to capture momentum diffusion and analogous to the role of the diffusion coefficient in Fick's second law for Brownian motion. The kinematic viscosity $\nu$ and diffusion coefficient are both proportional to $\ell^2/\tau$, where $\ell$ is mean free path and $\tau$ is the particle lifetime. However, as we discuss in Sec. I, the kinematic viscosity is not as easy to measure experimentally as the dynamic viscosity $\eta$. Having obtained the kinematic viscosity from electron-electron lifetime, the question that we need to address now is how do we convert it to dynamic viscosity. For classical fluid or 2D electron gas, the relation between them is given by $\eta = mn\nu$. This combination of $mn$ appears in the corresponding Navier-Stokes equation $mn\left(\partial_t + \mathbf{u}\cdot\nabla\right)\mathbf{u} = -en\mathbf{E} + \eta\nabla^2\mathbf{u} + (\eta/2)\nabla(\nabla\cdot\boldsymbol{u}) - \nabla P - mn\mathbf{u}/\tau_{\mathrm{dis}}$. In this work, we generalize the relation between dynamic viscosity and kinematic viscosity by deriving the corresponding Navier-Stokes equation that is valid for 2DEGs (parabolic bands), monolayer and bilayer graphene by multiplying the Boltzmann equation by momentum and then integrating over both momentum and the band index (see Appendix. B for details). The resulting Navier-Stokes equation is

$$\begin{aligned} \rho_m\left(\partial_t + \mathbf{u}\cdot\nabla\right)\mathbf{u} &= -en\mathbf{E} + \eta\nabla^2\mathbf{u} + \frac{\eta}{2}\nabla\left(\nabla\cdot\boldsymbol{u}\right) - \nabla P \\ &\quad - \left[\partial_t\rho_m + \nabla\cdot(\rho_m\mathbf{u})\right]\mathbf{u} - \rho_m\frac{\mathbf{u}}{\tau_{\mathrm{dis}}}. \end{aligned} \tag{4}$$

The Navier-Stokes equations that we have just derived allows us to infer that the dynamic viscosity is proportional to the kinematic viscosity by a proportionality constant that we call 'effective mass density', i. e. $\eta = \rho_m\nu$. The effective mass density $\rho_m$ for isotropic dispersion is given by

$$\rho_m = \sum_{\lambda=e,h} g \int \frac{d^2k}{(2\pi)^2} \frac{\hbar^2k^2}{2}\left(-\frac{\partial f^0_{\boldsymbol{k},\lambda}}{\partial\varepsilon_{\boldsymbol{k},\lambda}}\right). \tag{5}$$

where we have used the convention of $\varepsilon_{\boldsymbol{k},e(h)} = \varepsilon_{\boldsymbol{k},+}$.

By performing integration by parts, we can show that in monolayer graphene, the effective mass density $\rho_m$ is equivalent to $w/v_F^2$, where $w$ is the enthalpy density [57]. In this work, we find that the corresponding quantity $\rho_m$ for bilayer graphene is equivalent to $(n_e + n_h)m^\star$, where $n_e$ and $n_h$ are the electron and hole carrier density, respectively. This quantity $\rho_m$ reduces to $n_e m^\star$ for 2D electron gas and classical fluid as it should be.

Here, the momentum relaxation time $\tau_{\rm dis}$ is not experimentally observable, but is related to the Ohmic resistivity $\rho_{\rm dis}$ through $\rho_m/\tau_{\rm dis} = e^2(n_e^2\rho_{\rm dis}^{(e)} + n_h^2\rho_{\rm dis}^{(h)})$. The fifth term in the right hand side of the Navier-Stokes equation contains the term which resembles the continuity equation for particle density, except that it is now for the effective mass density. For 2D parabolic system that this term will vanish, but it is present for both monolayer and bilayer graphene. The temperature dependence of effective mass density emerge due to presence of electron and hole plasma in monolayer and bilayer graphene. The effective mass density for both monolayer and bilayer can be computed analytically (the results for the enthalpy density of monolayer graphene were previously obtained in Ref. [57])

$$\rho_m = -\frac{6}{\pi}\frac{(k_B T)^3}{\hbar^2 v_F^4}\sum_{\lambda=\pm1}\mathrm{Li}_3\left[-e^{\lambda\mu/(k_B T)}\right]; \quad \text{monolayer}, \tag{6a}$$

$$\rho_m = \frac{2}{\pi}\frac{(m^\star)^2 k_B T}{\hbar^2}\sum_{\lambda=\pm1}\ln\left[1+e^{\lambda\mu/(k_B T)}\right]; \quad \text{bilayer}, \tag{6b}$$

where $\mathrm{Li}_3$ is trilogarithm function, defined by $\mathrm{Li}_3(x) = \sum_{n=1}^{\infty} x^n/(n^3)$. At low density or high temperature ($T \gg T_F$), we find that the effective mass density of monolayer and bilayer graphene goes as

$$\rho_m(T \gg T_F) \approx \frac{9\zeta(3)}{\pi}\frac{(k_B T)^3}{\hbar^2 v_F^4}; \qquad \text{monolayer}, \tag{7a}$$

$$\rho_m(T \gg T_F) \approx \frac{4\ln 2}{\pi}\frac{(m^\star)^2 k_B T}{\hbar^2}; \qquad \text{bilayer}, \tag{7b}$$

while in the opposite limit ($T \ll T_F$), it is given by

$$\rho_m(T \ll T_F) \approx \frac{1}{\pi}\frac{|E_F|^3}{\hbar^2 v_F^4}\left[1+\frac{\pi^2}{2}\left(\frac{T}{T_F}\right)^2\right]; \qquad \text{monolayer}, \tag{8a}$$

$$\rho_m(T \ll T_F) \approx \frac{2}{\pi}\frac{(m^\star)^2|\varepsilon_F|}{\hbar^2} = |n|m^\star; \qquad \text{bilayer}. \tag{8b}$$

We note that the low temperature limit of effective mass density of bilayer graphene is similar to the 2D electron gas and classical fluid [56].

## III. TEMPERATURE AND COULOMB INTERACTION STRENGTH DEPENDENCE OF DYNAMIC VISCOSITY OF MONOLAYER AND BILAYER GRAPHENE

Having obtained the effective mass density $\rho_m$, we can now proceed to calculate the dynamic viscosity by combining it with the kinematic viscosity obtained from the lifetime $\tau_{ee}$. Our results for the dynamic viscosity for both monolayer and bilayer graphene are shown in Fig. 1a and b, respectively. This result is to be understood in the context of when the hydrodynamic regime holds, i. e. when electron-electron lifetime is less than electron-impurity and electron-phonon lifetime, as well as timescale $W/v_F$ where $W$ is sample dimension, as laid out in our previous work on the determination of hydrodynamic window in monolayer and bilayer graphene from scattering rates [50]. In the case where it lies outside the hydrodynamic window, we should regard it as a theoretical ideal limit.

In the degenerate limit ($T \ll T_F$), we obtain $E_F\tau_{ee}/\hbar(T \ll T_F) \propto (T_F/T)^2$ for both monolayer and bilayer from our numerical calculations. Combining with low temperature asymptotes for effective mass density and effective velocity we obtained earlier in section II, we get

$$\eta\,(T \ll T_F) \propto \frac{1}{\hbar v_F^2}\frac{\varepsilon_F^4}{(k_B T)^2}; \qquad \text{monolayer}, \tag{9a}$$

$$\eta\,(T \ll T_F) \propto \frac{m^\star}{\hbar}\frac{|\varepsilon_F|^3}{(k_B T)^2}; \qquad \text{bilayer}. \tag{9b}$$

The temperature dependence of the dynamic viscosity of both monolayer and bilayer at $T \ll T_F$ goes as $\eta \propto T^{-2}$, which is similar to the 2D electron gas with a parabolic dispersion and the classical fluid. This temperature scaling of viscosity is related to the well known $T^2$ resistivity that arises from multiband band systems with different electron masses [58] or umklapp scattering [59, 60]. This temperature dependence comes solely from the lifetime of both monolayer and bilayer graphene itself which goes as $\tau_{ee}(T \ll T_F) \propto T^{-2}$ as their effective mass density are nearly independent of temperature in the low temperature regime. The high temperature regime ($T \gg T_F$) is more interesting. Our numerical calculations reveal that $E_F\tau_{ee}/\hbar(T \gg T_F) \propto T_F/T$ for both monolayer and bilayer. Combining this with high temperature asymptotes for effective mass density and effective velocity we obtained earlier in section II, we observe that the dynamic viscosity becomes density independent. In particular, we find

$$\eta\,(T \gg T_F) = F_m(\alpha)\frac{(k_B T)^2}{\hbar v_F^2}; \qquad \text{monolayer}, \tag{10a}$$

$$\eta\,(T \gg T_F) = F_b(\alpha)\frac{m^\star k_B T}{\hbar}; \qquad \text{bilayer}, \tag{10b}$$

where $F_{m,b}(\alpha)$ is a one parameter function which depends only on the electron-electron interaction strength.

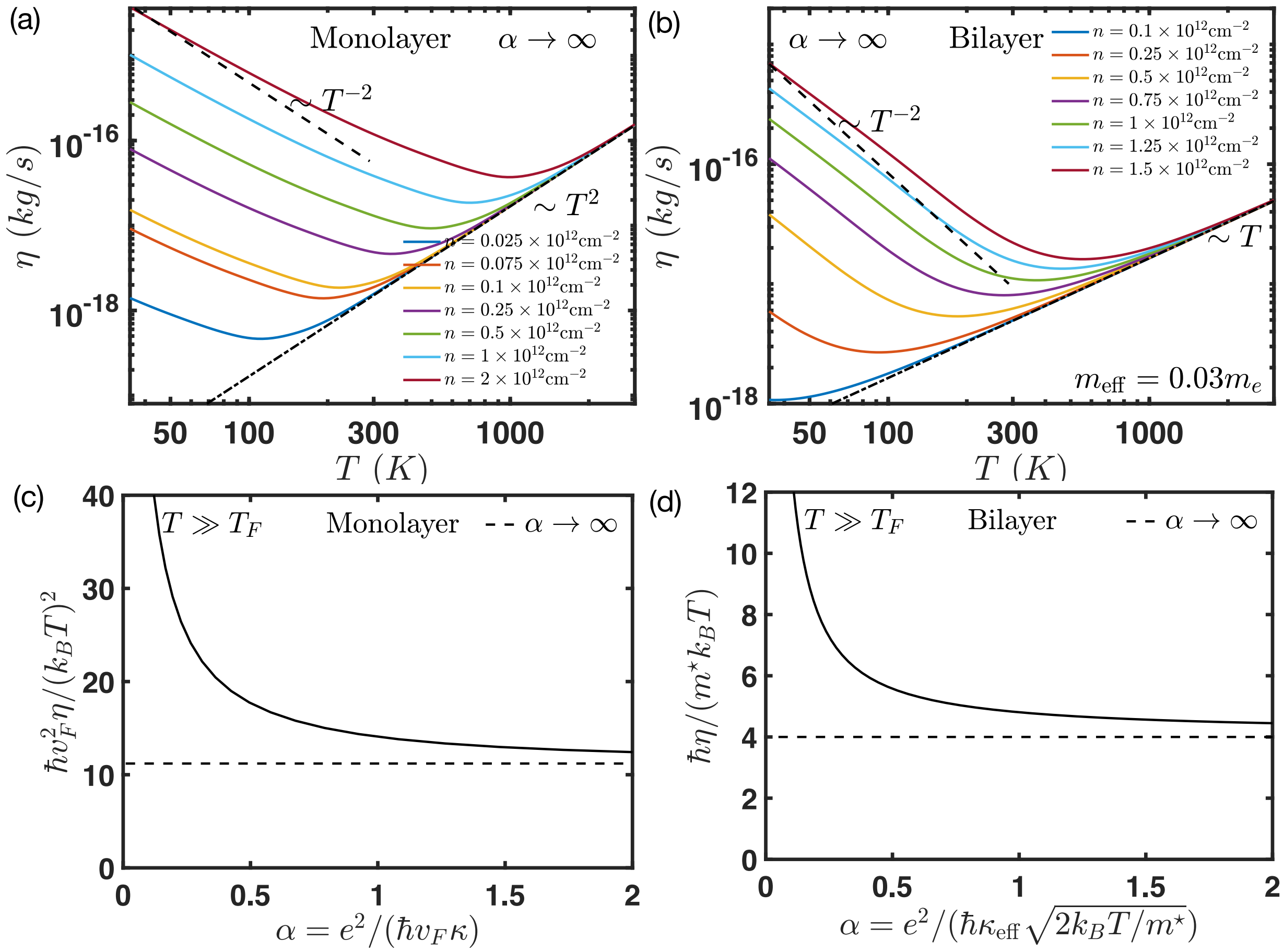

Figure 1. Dynamical hydrodynamic viscosity as a function of temperature and Coulomb interaction strength. Top panels show that for both monolayer and bilayer graphene, the viscosity is non-monotonic with temperature. At low temperature, the degenerate regime for unipolar hydrodynamics gives a density-dependent viscosity that increases at low temperature in contrast to high-temperature, where the ambipolar transport becomes density independent. Bottom panels show that with increasing Coulomb interactions, the viscosity approaches a universal limit that is independent of both carrier density and interaction strength.

In this limit, dynamic viscosity increases with increasing temperature. We can understand this since in the high temperature regime, the electron-electron lifetime has a weaker decay ($\propto T^{-1}$) compared to the rapid increase of effective mass density ($w/v_F^2 \propto T^3$ for MLG and $(n_e + n_h)m^\star \propto T$ for BLG). More interestingly, we find that for sufficiently strong Coulomb interaction ($\alpha \to \infty$), the dynamic viscosity is also independent of electron-electron interaction strength, $F_m(\alpha \to \infty) \approx 11.2$ and $F_b(\alpha \to \infty) \approx 4$ for monolayer and bilayer graphene, respectively. The dependence of the dynamic viscosity (scaled by $T^2$ for monolayer and by $T$ for bilayer) with electron-electron interaction strength is shown in Fig. 1c and d. We have characterized the electron-electron interaction strength by $\alpha = e^2/(\hbar v_F \kappa_{\mathrm{eff}})$ for monolayer and $\alpha = e^2/(\hbar \kappa_{\mathrm{eff}} \sqrt{2k_B T/m^\star})$ for bilayer graphene.

In Fig. 2a and b, we show the dynamic viscosity to entropy density ratio in the limit for strong Coulomb interaction ($\alpha \to \infty$) for monolayer and bilayer graphene, respectively. While the entropy density of monolayer has been calculated previously (see e. g. Ref. [57]), which is given by

$$s = -\frac{1}{T}\left\{\frac{6}{\pi}\frac{(k_B T)^3}{(\hbar v_F)^2}\sum_{\lambda=\pm 1} \mathrm{Li}_3\left[-\exp\left(\lambda \frac{\mu}{k_B T}\right)\right] + \mu n\right\}, \tag{11}$$

($\mathrm{Li}_3$ is trilogarithm function as defined in section. II), the entropy density for bilayer is calculated for the first time in this work.

The entropy density for bilayer graphene can be obtained from Euler relation

$$\varepsilon = Ts - P + \mu n, \tag{12}$$

where $P$ is pressure and $\mu$ is chemical potential which for bilayer graphene is equal to the Fermi energy, i. e. $\mu = E_F$

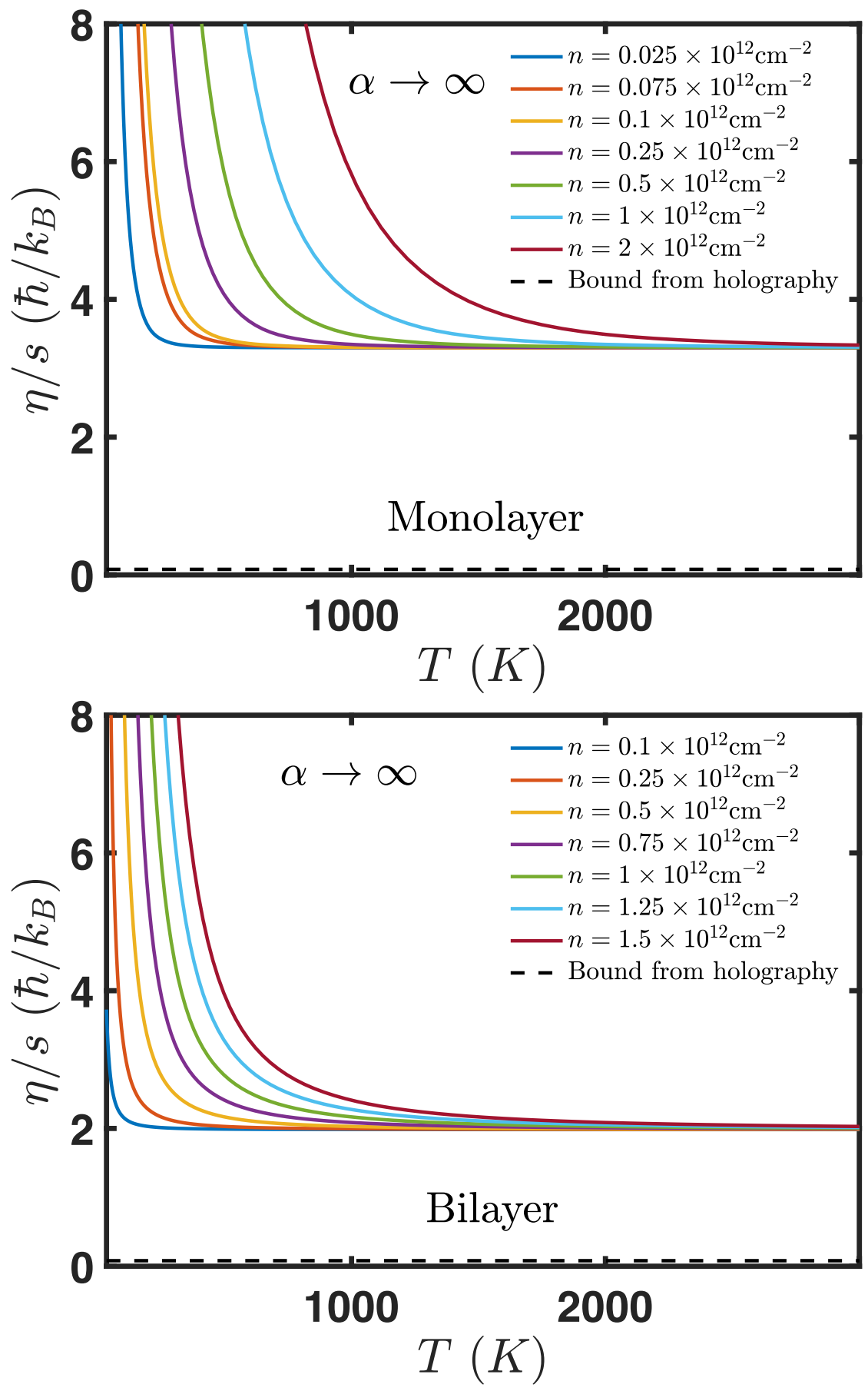

Figure 2. Dynamic viscosity by entropy density of (a) Monolayer graphene (b) Bilayer graphene. At high temperature or low density ($T \gg T_F$) the dynamic viscosity by entropy density ratio of both monolayer and bilayer graphene reach the universal value of 0.8 and 0.5 ($\hbar/k_B$) respectively.

and $\varepsilon$ is internal energy density, given by

$$\begin{aligned}\varepsilon &= \sum_{\lambda=e(h)} g \int \frac{d^2k}{(2\pi)^2} \varepsilon_{\mathbf{k}} f^{(0)}_{\mathbf{k},\lambda} \\ &= -\frac{2}{\pi}\frac{m^\star (k_B T)^2}{\hbar^2} \sum_{\lambda=\pm 1} \mathrm{Li}_2\left[-\exp\left(\lambda \frac{\varepsilon_F}{k_B T}\right)\right], \end{aligned} \tag{13}$$

where $\mathrm{Li}_2(z) = \int_z^0 dx\, x^{-1} \ln(1-x)$ is the dilogarithm function.

From the partition function of electron(holes)

$$\mathcal{Z}_{e(h)} = \prod_{\mathbf{k}} \left[1 + e^{-\beta(\varepsilon_{\mathbf{k}} \mp \mu)}\right]^g, \tag{14}$$

we can obtain the grand potential $\Omega$ by

$$\Omega = -k_B T \sum_{\lambda=e,h} \ln \mathcal{Z}_\lambda, \tag{15}$$

which leads to expression for pressure by $\Omega = -PA$

$$P = -\frac{2}{\pi}\frac{m^\star (k_B T)^2}{\hbar^2} \sum_{\lambda=\pm 1} \mathrm{Li}_2\left[-\exp\left(\lambda \frac{\varepsilon_F}{k_B T}\right)\right], \tag{16}$$

which is similar to the internal energy density, i. e. $P = \varepsilon$.

Finally, the bilayer graphene entropy density is given by

$$s = -\frac{4m^\star k_B^2 T}{\pi\hbar^2} \sum_{\lambda=\pm 1} \left\{ \mathrm{Li}_2\left[-\exp\left(\lambda \frac{\varepsilon_F}{k_B T}\right)\right]\right\} - \frac{2m^\star \varepsilon_F^2}{\pi \hbar^2 T}. \tag{17}$$

The high temperature or low density limit of the entropy density for both monolayer and bilayer is given by

$$s(T \gg T_F) \approx \frac{9\zeta(3)}{\pi}\frac{k_B^3 T^2}{\hbar^2 v_F^2}; \qquad \text{monolayer}, \tag{18a}$$

$$s(T \gg T_F) \approx \frac{2\pi}{3}\frac{m^\star k_B^2 T}{\hbar^2}; \qquad \text{bilayer}. \tag{18b}$$

Since both the entropy density and the dynamic viscosity have similar leading order temperature dependence at $T \gg T_F$, their ratio becomes universal, i. e. independent of temperature. This asymptote is given by $\eta/s = C_{\mathrm{mono(bi)}}\hbar/k_B$, where $C_{\mathrm{mono}} \approx 3.2$ for monolayer graphene and $C_{\mathrm{bi}} \approx 2$ for bilayer graphene. We observe that the dynamic viscosity to entropy density ratio of bilayer graphene is lower than that of monolayer and closer to the theoretical bound calculated from the holographic principle [26]. Noticing the trend that the higher power of the momentum dependence of the energy dispersion gives lower entropy density to viscosity ratio, we conjectured that a two carrier system with energy dispersion of $\varepsilon_F \propto \pm k_F^p$, would have closer entropy density to viscosity ratio to the theoretical bound as the power $p$ increases.

## IV. STABILITY OF THE UNIVERSAL FLUID IN BILAYER GRAPHENE

Though we have calculated the dynamic viscosity in bilayer graphene in the theoretical limit of strong Coulomb interaction, the possibility of formation of competing order in this strong coupling limit needs to be examined carefully. In this section, we take a more careful look into the possibility of quantum phase transition into exciton condensates in a more realistic condition due to increasing strength of Coulomb interactions.

First, we briefly explain different categories of excitons based on the nature of their interaction-driven quantum phase transition [61]. The first category is Class I, which corresponds to semi-metal to exciton condensate phase transition and the latter is Class II, which corresponds to band insulator to exciton condensate phase transition. Since opening of a small band gap is generic for bilayer graphene, we focus here on Class II excitons. The opening of a band gap in bilayer graphene is attributed to the asymmetry of on-site energies in each of the layers and was anticipated theoretically [52] and observed experimentally [15, 62]. This asymmetry could be due to various reasons, for instance due to externally applied gating. The external gating will adjust the carrier density and simultaneously open a band gap. Self-consistent Hartree calculation of band gap opening as a function of charge density in bilayer graphene has been reported in Ref. [63]. The authors found that the addition of density of $n \sim 10^{12}$ cm$^{-2}$ yields a band gap of $\Delta \sim 10$ meV. Once the band gap opens, bilayer graphene could be driven to Class II exciton condensates (exciton insulators) by increasing the strength of electron-electron interactions until it exceeds a critical value. To estimate this, we first calculate the exciton binding energy by solving the Lippmann-Schwinger equation of gapped bilayer graphene [64] with energy dispersion $E_{\boldsymbol{k},\pm} = \pm\sqrt{(\hbar^2 k^2/2m^\star)^2 + (\Delta/2)^2}$ , i.e.

$$(\Delta - E_b - 2E_{\boldsymbol{k}})\,\Psi_{\boldsymbol{k}} = \int \frac{d^2k'}{(2\pi)^2}\frac{V\left(\boldsymbol{k}-\boldsymbol{k}'\right)}{\varepsilon\left(\boldsymbol{k}-\boldsymbol{k}'\right)} Z_{\tau,\tau'}\left(\boldsymbol{k},\boldsymbol{k}'\right)\Psi_{\boldsymbol{k}'}, \tag{19}$$

Here $\Psi_{\boldsymbol{k}}$ is the exciton wave function, $E_b$ is the exciton binding energy, $E_{\boldsymbol{k}} = |E_{\boldsymbol{k},\pm}|$, $\Delta$ is band gap and $Z_{\tau,\tau'}\left(\boldsymbol{k},\boldsymbol{k}'\right) = \langle\phi^{(-)}_{\boldsymbol{k},\tau}|\phi^{(-)}_{\boldsymbol{k}',\tau}\rangle\langle\phi^{(+)}_{\boldsymbol{k}',\tau'}|\phi^{(+)}_{\boldsymbol{k},\tau'}\rangle$ is the vertex form factors where $|\phi^{(\lambda)}_{\boldsymbol{k},\tau}\rangle$ is the eigenstates of the gapped bilayer graphene bare hamiltonian with $\tau$ is valley index and $\lambda$ is band index. We use the static RPA polarizability $\Pi(\boldsymbol{q})$ to screen the bare Coulomb potential $V = -2\pi e^2/\kappa q$ via dielectric function $\varepsilon(\boldsymbol{q}) = 1 - V(\boldsymbol{q})\Pi(\boldsymbol{q})$. The excitonic states $\Psi_{\boldsymbol{k}}$ could be decomposed into different angular momentum channel $l$ by $\Psi_{\boldsymbol{k}} = \sum_\ell \sqrt{k_0/k}\,\psi^\ell_k e^{i\ell\theta_k}$ where $k_0 = \sqrt{m^\star\Delta}/\hbar$.

Remarkably, we find that the dimensionless exciton binding energy $E_b/\Delta$ is given by a one-parameter function $F_{\ell,\tau\tau'}[e^2/(\hbar\kappa\sqrt{\Delta/m^\star})]$ (see inset of Fig. 3). We find numerically that the scaling function obeys the following asymptotes: $F_{\ell,\tau\tau'}(x \ll 1) \sim A_{\ell,\tau\tau'}x^{m_{\ell,\tau\tau'}}$ with $A_{0,+} \approx 1$ and $m_{0,+} \approx 1.3$ ($A_{1,+} \approx 0.3$ and $m_{1,+} \approx 1.3$) for $s$-wave ($p$-wave) intravalley exciton, while in the opposite limit it approaches a constant value, $F_{\ell,\tau\tau'}(x \to \infty) \sim C^{(\infty)}_{\ell,\tau\tau'}$ with $C^{(\infty)}_{0,+} \approx 3.9$ ($C^{(\infty)}_{1,+} \approx 3.6$) for $s$-wave ($p$-wave) intravalley exciton. Here, we only focus on the intravalley exciton since the exciton binding energy difference between intravalley and intervalley exciton is negligible, with less than 1% difference [64].

The critical electron-electron interactions strength for exciton formation is finally obtained from solving $E_b = \Delta$. We find that the critical electron-electron interactions strength for exciton formation $[e^2/(\hbar\kappa\sqrt{\Delta/m^\star})]_{\rm crit}$ equals to a universal value that only depends on the angular momentum channel $\ell$ and the nature of electron-hole pairing (intravalley or intervalley exciton), i. e.

$$\left(\frac{e^2}{\hbar\kappa\sqrt{\Delta/m^\star}}\right)_{\rm crit} = C_{\ell,\tau\tau'}. \tag{20}$$

We find numerically that $C_{0,+} \approx 4.5$ ($C_{1,+} \approx 7.9$) for $s$-wave ($p$-wave) intravalley excitons. Combining this with the definition of dimensionless Coulomb interaction strength for bilayer graphene $\alpha$ found earlier in Sec. III, the condition for $e^2/(\hbar\kappa\sqrt{\Delta/m^\star})_{\rm crit}$ translates into a condition for $\alpha_{\rm crit}$ given by

$$\alpha_{\rm crit} = C_{\ell,\tau\tau'}\sqrt{\frac{\Delta}{2k_BT}}. \tag{21}$$

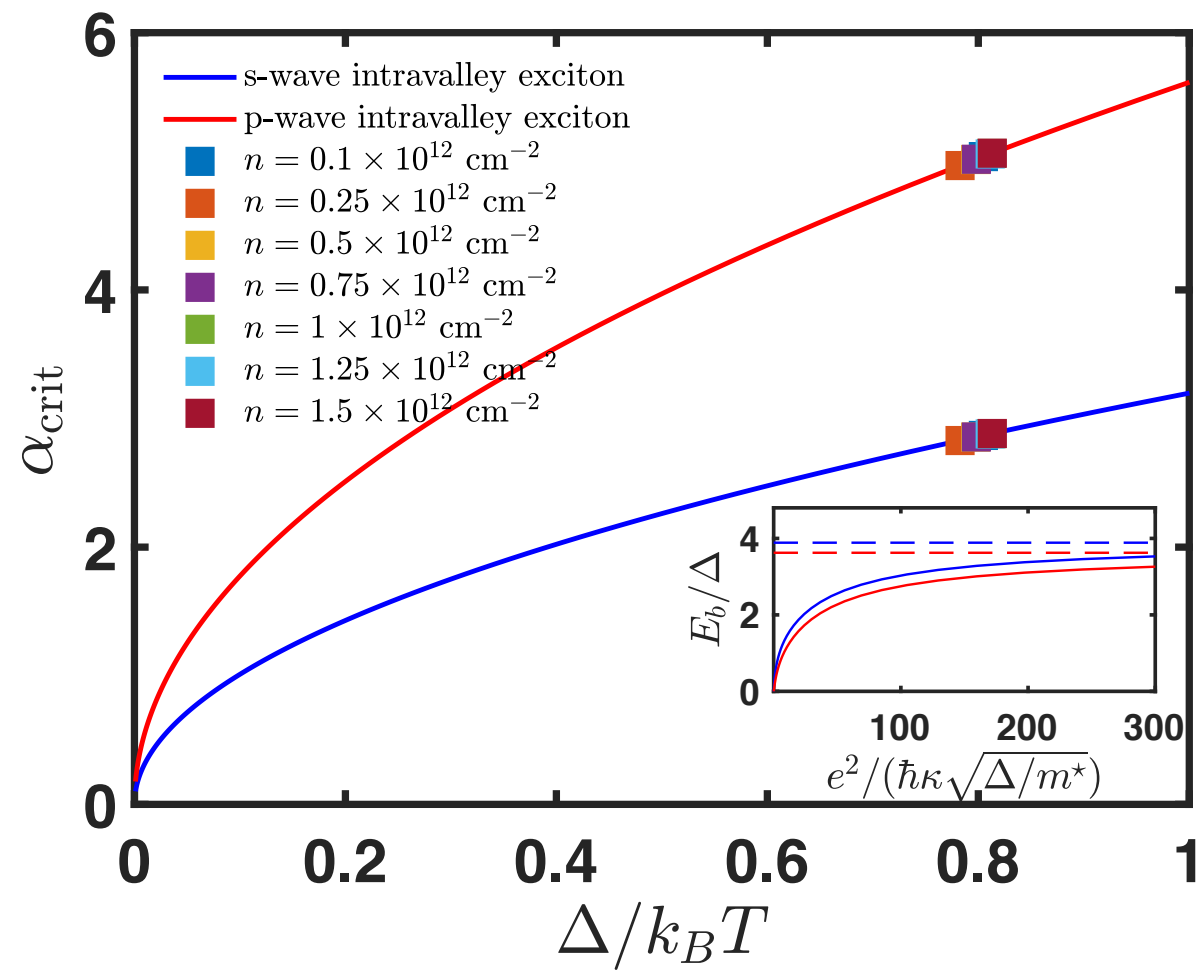

Figure 3. Critical electron interaction strength for exciton formation in bilayer graphene increases with band gap over temperature $\Delta/k_BT$. Square marker corresponds to density and temperature $(n, T^\star[n])$ where $T^\star$ is the crossover temperature from degenerate to plasma regime, corresponding to minima in the temperature dependence of viscosity. Inset shows a universal function of dimensionless exciton binding energy $E_b/\Delta$ against $e^2/(\hbar\kappa\sqrt{\Delta/m^\star})$. In the limit of very strong electron-electron interaction, the dimensionless binding energy $E_b/\Delta$ saturates to a constant value. Blue line denotes $s$-wave and red line denotes $p$-wave intravalley exciton solution. Dashed lines are asymptotes for $E_b/\Delta$ as $e^2/(\hbar\kappa\sqrt{\Delta/m^\star}) \to \infty$.

As seen in Fig. 3, the critical electron-electron interactions strength increases with $\Delta/k_BT$. We also show the critical Coulomb interaction strength corresponding to the density and temperature $(n, T^\star[n])$ where each density $n$ is similar to what we have considered earlier in Fig. 1b, and $T^\star[n]$ is the crossover temperature from degenerate to plasma regime, which corresponds to the minima of viscosity in the temperature dependence dynamic viscosity in Fig. 1b. We have calculated the band gap $\Delta[n]$ using the self-consistent Hartree calculation of band gap opening [63]. We can see that for these particular case, the quasiparticle is stable as long as $\alpha \lesssim 3$ ($\alpha \lesssim 5$) for $s$-wave ($p$-wave) intravalley exciton. Typical values of Coulomb interaction strength e. g. for BLG/hBN is $\alpha \approx 1.25 - 3$ at temperature $T \approx 50 - 300$ K. Since $\alpha \lesssim \alpha_{\rm crit}$ for the experimentally relevant regime (clustered squares in the figure), it should be possible to observe the non-monotonic viscosity experimentally. Moreover, one can read off from Fig. 3 the value of $\alpha_{\rm crit}$ for which the universal hydrodynamic fluid breaks down into an exciton fluid.

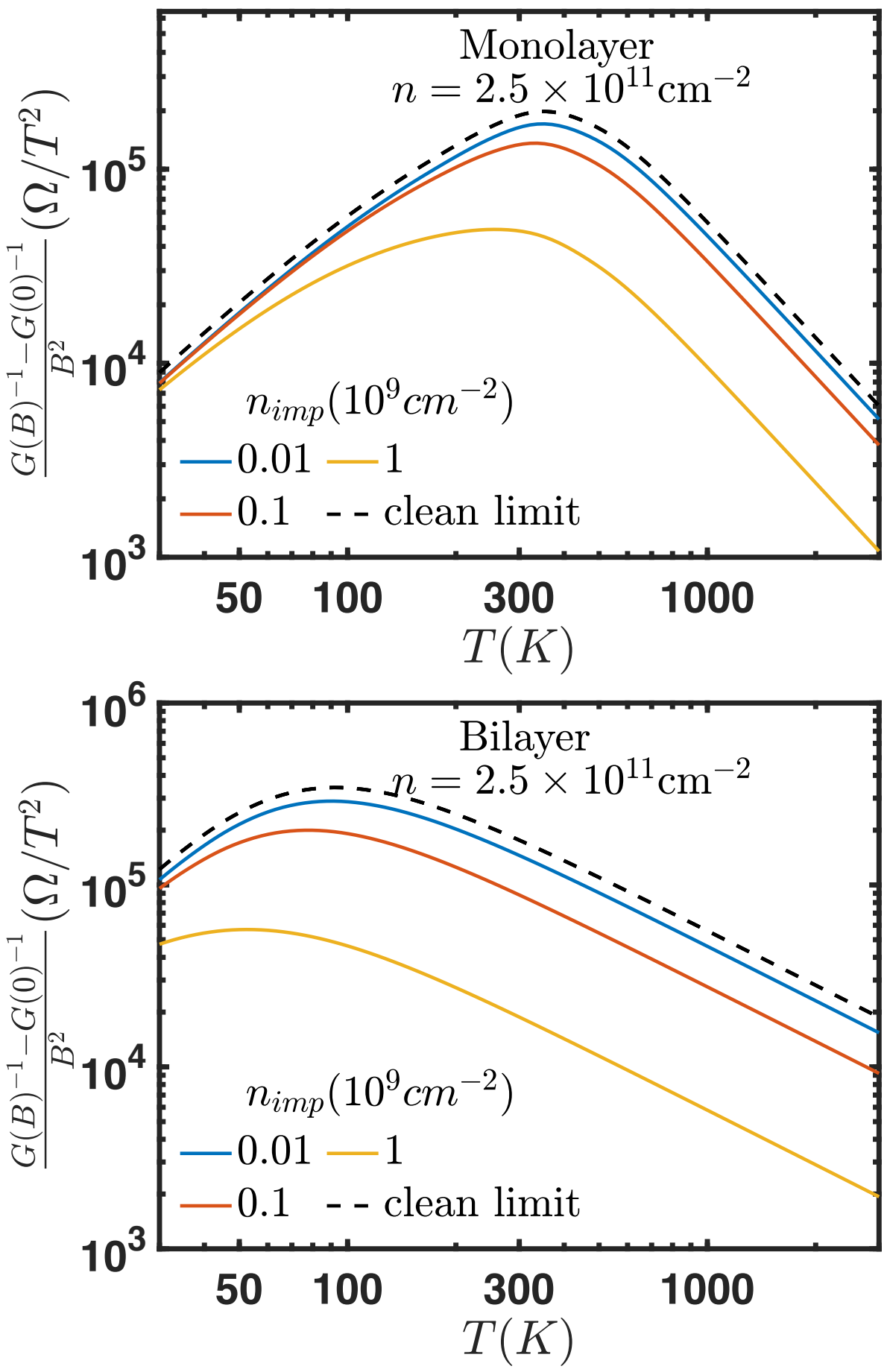

Figure 4. Coefficient of $B^2$ of inverse magneto conductance with temperature in the Corbino. The non-monotonic behaviour is reflected in the magneto-conductance in the clean limit (dashed lines). Increasing impurity concentration reduces this effect.

## V. MAGNETOCONDUCTANCE IN CORBINO GEOMETRY

Finally, we look into the experimental signature of dynamic viscosity by considering inverse magnetoconductivity in a Corbino ring. Unlike a Hall bar, the Corbino geometry has a distinct dependence of magnetoconductance on magnetic field $B$ which allows us to determine viscosity. The magneto-conductance in the clean limit was previously calculated by Ref. [31], assuming the Fermi-liquid regime. Here, we generalise this result to the viscous plasma regime by using the dynamic viscosity calculated earlier. Our results for the magnetoconductance as function of carrier density and temperature are shown in Fig. 4. It should be noted that only the magneto-conductance can be measured in the Corbino because of vanishing Hall voltage but a non-vanishing Hall current. We contrast this to a more realistic case of viscous flow with disorder scattering rate $1/\tau_{\rm dis}$. The magneto-conductance [32] is

$$G(B)^{-1} = \frac{\rho_{\rm dis}}{2\pi}\ln\left(\frac{r_2}{r_1}\right) + \frac{B^2\tau_{\rm dis}}{2\pi\rho_m} \times \left[c_1 I_0(r/l_G) + c_2 K_0(r/l_G) + \ln(r/l_G)\right]_{r_1}^{r_2} \tag{22}$$

where $l_G = \sqrt{\eta\tau_{\rm dis}/\rho_m}$ is the Gurzhi length, $I_0, K_0$ stand for the zeroth order modified Bessel functions. $c_1, c_2$ are coefficients to be determined from the boundary conditions of flow at $r_1, r_2$ (assumed to be no-slip in figure). Equation 22 holds for $B$ sufficiently small ($2\omega_c\tau_{ee} \ll 1$) so that quantum effects as well as Hall viscosity are negligible. It can be confirmed that in the limit of low impurity scattering ($\tau_{\rm dis}^{-1} \to 0$), this expression reduces to the impurity-free magnetoresistance derived previously in Ref. [31]. We emphasize that the momentum relaxation time $\tau_{\rm dis}$ is not experimentally observable. Rather, the combination $\tau_{\rm dis}/\rho_m$ can be obtained from the $B = 0$ transport and the Gurzhi length $l_G$ measured from the magnetoconductance (see Eq. 22). It is therefore the dynamic viscosity $\eta$ and not the kinematic viscosity $\eta/\rho_m$ that is the relevant quantity measured experimentally. More explicitly, the ratio $\tau_{\rm dis}/\rho_m = e(n_e/\mu^e_{\rm dis} + n_h/\mu^h_{\rm dis})$ requires the determination of both the electron mobility $\mu^e_{\rm dis}$ and hole mobility $\mu^h_{\rm dis}$ separately. However, this expression simplifies for both $T \gg T_F$, where $\tau_{\rm dis}/\rho_m \approx 2e^2(n_e+n_h)^2\rho_{\rm dis}$, and for $T \ll T_F$, where $\tau_{\rm dis}/\rho_m \approx e^2n^2\rho_{\rm dis}$.

## VI. CONCLUSION

Electron hydrodynamics is a new and interesting regime for electronic transport in condensed matter systems. In this regime, a new transport behaviour that is characterized by a phenomenological dynamical viscosity in the Navier-Stokes equation emerges. To better understand this phenomena, we have performed a microscopic calculations of the dynamic viscosity for both monolayer and bilayer graphene within the random phase approximation and found that it is non-monotonic with temperature. At low temperature, the degenerate unipolar carriers generate a dynamic viscosity that decreases with temperature as ($\sim 1/T^2$) as one might expect for a classical fluid. However, the dynamical viscosity then increases with temperature as $\sim T^2$ for monolayer graphene and $\sim T$ for bilayer graphene at high temperature. We understand this different behavior as the interplay between the temperature dependence of the electron-electron lifetime and the effective mass density. The high-temperature regime is universal in that the dynamical viscosity becomes independent of both carrier density and interaction strength. For strong interactions, bilayer graphene more closely approaches the holographic bound compared to monolayer graphene. Although this strongly correlated electron fluid is unstable to exciton formation, we show that for realistic experimental parameters, bilayer graphene is in the hydrodynamic regime. Finally, we discuss how the dynamical viscosity can be measured experimentally using magnetoconductance measurements in a Corbino geometry. Having a quantum hydrodynamic fluid with tunable viscosity opens the door to exploring new physics and new devices that exploit the hydrodynamic regime of electrons.

## ACKNOWLEDGMENTS

It is a pleasure to thank Cory Dean for sharing his unpublished data and for valuable discussions. We also thank Aydin Keser, Oleg Sushkov and Giovanni Vignale for collaboration on related projects, and for discussions. We acknowledge the financial support from Singapore National Research Foundation Investigator Award (NRF-NRFI06-2020-0003) and the of use dedicated research computing resources at CA2DM.

### Appendix A: Viscosity transport time and quasiparticle lifetime

We calculate the viscosity transport time for monolayer following Ref. [42]. In general, one need to solve the self-consistent Bethe-Salpeter equation to obtain this. However, in the regime of $T \ll T_F$, a set of approximations allows us to dramatically simplify the expressions and in this limit it is given by

$$\frac{\hbar}{\tau_{ee,2}(\varepsilon)} = -2\int \frac{d^2\boldsymbol{q}}{(2\pi)^2}\sum_{\lambda'} \Im m\left[\frac{V(q)}{\epsilon(q,\omega-\varepsilon_{\boldsymbol{k}-\boldsymbol{q},\lambda'}+\mu,T)}\right] \mathcal{F}_{\lambda\lambda'}(\theta_{\boldsymbol{k},\boldsymbol{k}-\boldsymbol{q}})\sin^2\theta_{\boldsymbol{k},\boldsymbol{k}-\boldsymbol{q}}[n_{\rm B}(\omega-\varepsilon_{\boldsymbol{k}-\boldsymbol{q},\lambda'}+\mu)+n_{\rm F}(-\varepsilon_{\boldsymbol{k}-\boldsymbol{q},\lambda'}+\mu)] \quad (T \ll T_F), \tag{A1}$$

which looks similar to the the expression for quasiparticle lifetime, except for the factor of $\sin^2\theta_{\boldsymbol{k},\boldsymbol{k}-\boldsymbol{q}}$. Here $V(q) = 2\pi e^2/(\kappa q)$ is the Coulomb potential, $n_{\rm B/F}(x) \equiv 1/[\exp(\beta x) \mp 1]$ the Bose (Fermi) distribution function, $\lambda, \lambda'$ are band indices ($\pm$ denote the conduction and valence bands), $\mathcal{F}_{\lambda\lambda'}(\varphi)$ is the chirality factor, and $\varphi$ the scattering angle. $\epsilon(q,\omega,T) \equiv 1 - V(q)\Pi(q,\omega,T)$ refers to the dynamical RPA dielectric function.

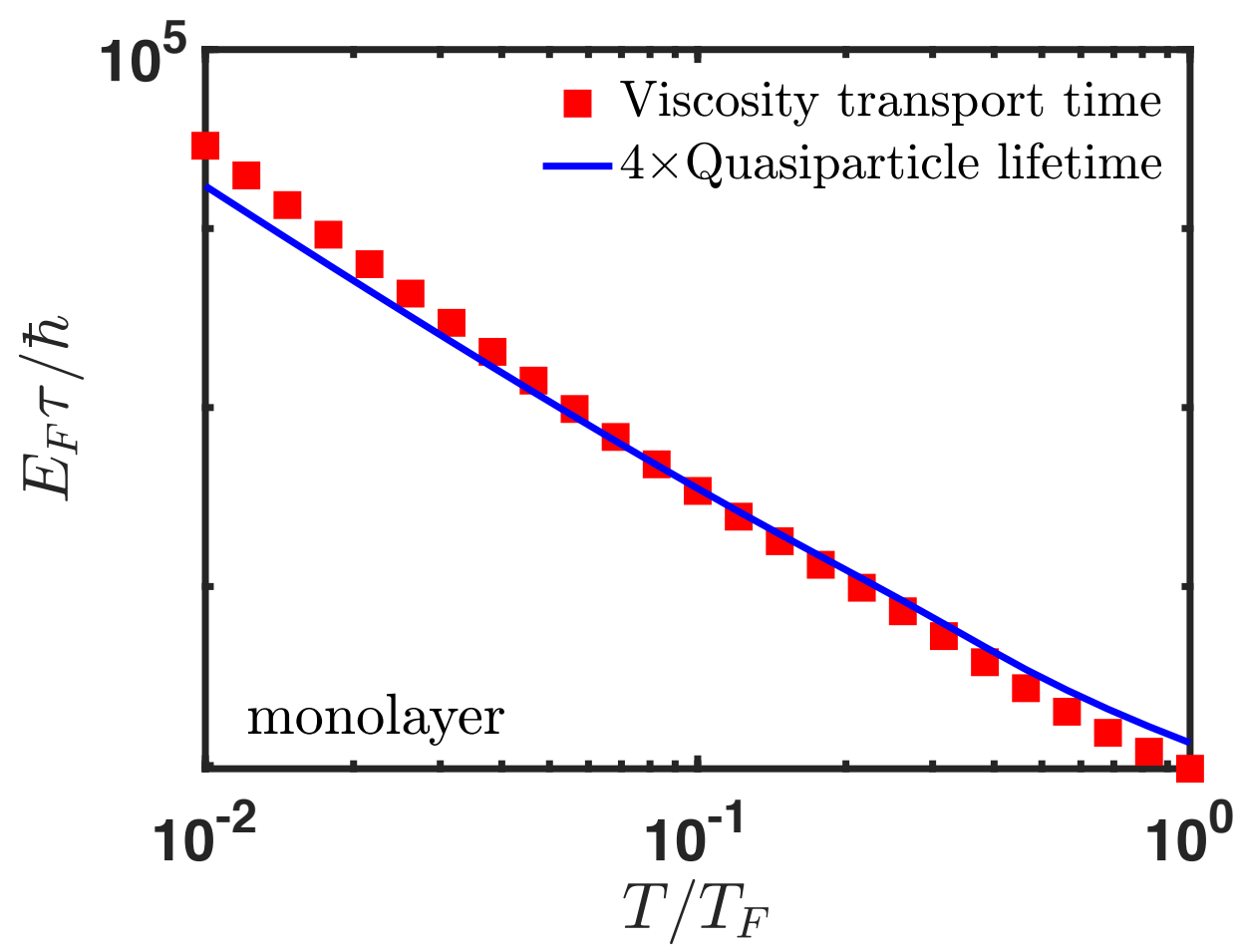

Figure A.1. Viscosity transport time can be very well approximated with 4×quasiparticle lifetime for $T \ll T_F$.

We found that the viscosity transport time can be very well approximated with 4 times quasiparticle lifetime as shown in figure A.1. We use this expressions throughout our work.

### Appendix B: Derivation of Navier-Stokes equation from Boltzmann equation

The quantum Boltzmann equation for electron(holes) in the presence of electric field is

$$\left(\partial_t + \boldsymbol{v}_{\boldsymbol{k},e(h)}\cdot\nabla_{\mathbf{r}} - \lambda e\boldsymbol{E}\cdot\frac{1}{\hbar}\nabla_{\boldsymbol{k}}\right) f_{\boldsymbol{k},e(h)} = \mathcal{I}^{C}_{\boldsymbol{k},e(h)}\left[\{h_{\boldsymbol{k}_i,\lambda_i}\}\right] + \mathcal{I}^{\rm dis}_{\boldsymbol{k},e(h)}\left[\{h_{\boldsymbol{k}_i,\lambda_i}\}\right], \tag{B1}$$

where $\boldsymbol{v}_{\boldsymbol{k},e(h)}$ is group velocity, defined by $\boldsymbol{v}_{\boldsymbol{k},e(h)} = (1/\hbar)\partial\varepsilon_{\boldsymbol{k},e(h)}/\partial k$, $f_{\boldsymbol{k},e(h)}$ is distribution function, $\mathcal{I}^{C}_{\boldsymbol{k},e(h)}\left[\{h_{\boldsymbol{k}_i,\lambda_i}\}\right]$ and $\mathcal{I}^{\rm dis}_{\boldsymbol{k},e(h)}\left[\{h_{\boldsymbol{k}_i,\lambda_i}\}\right]$ is linearized collision integral for Coulomb interaction and disorder, respectively. In order to linearize the collision integral, we have make use of ansatz $f_{\boldsymbol{k},e(h)} \approx f^{(0)}_{\boldsymbol{k},e(h)} + \left(-\partial f^{(0)}_{\boldsymbol{k},e(h)}/\partial\varepsilon_{\boldsymbol{k},e(h)}\right) h_{\boldsymbol{k},e(h)}$.

The Navier-Stokes equation for electron(holes) is obtained using from Boltzmann equation by multiplication by momentum $\hbar\mathbf{k}$ and performing integration $g\int\frac{d^2k}{(2\pi)^2}$

$$\partial_t\left[g\int\frac{d^2k}{(2\pi)^2}\hbar\boldsymbol{k}\left(-\frac{\partial f^{(0)}_{\boldsymbol{k},e(h)}}{\partial E_{\boldsymbol{k},e(h)}}\right)\hbar\boldsymbol{k}\cdot\boldsymbol{u}_{e(h)}\left(\boldsymbol{r},t\right)\right]+\nabla_\alpha\left[g\int\frac{d^2k}{(2\pi)^2}\hbar\boldsymbol{k}v^\alpha_{\boldsymbol{k},e(h)}f_{\mathbf{k},e(h)}\left(\boldsymbol{r},t\right)\right]$$
$$+\lambda eg\int\frac{d^2k}{(2\pi)^2}\hbar\boldsymbol{k}\left(-\frac{\partial f^{(0)}_{\boldsymbol{k},e(h)}}{\partial E_{\boldsymbol{k},e(h)}}\right)\boldsymbol{v}_{\boldsymbol{k},e(h)}\cdot\boldsymbol{E}=g\int\frac{d^2k}{(2\pi)^2}\hbar\boldsymbol{k}\mathcal{I}_{\boldsymbol{k},e(h)}\left[\{h_{\boldsymbol{k}_i,\lambda_i}\}\right],\tag{B2}$$

Here we have used ansatz $h_{\boldsymbol{k},e(h)}=\hbar\mathbf{k}\cdot\boldsymbol{u}_{e(h)}$.

Assuming isotropic energy dispersion, we define/recognize

$$\rho_{m,e(h)}=g\int\frac{d^2k}{(2\pi)^2}\frac{(\hbar k)^2}{2}\left(-\frac{\partial f^{(0)}_{\mathbf{k},e(h)}}{\partial E_{\mathbf{k},e(h)}}\right)\tag{B3}$$

$$\Pi^{\alpha\beta}_{e(h)}=g\int\frac{d^2k}{(2\pi)^2}\hbar v^\alpha_{\boldsymbol{k},e(h)}k_\beta f_{\mathbf{k},e(h)}\left(\boldsymbol{r},t\right)\tag{B4}$$

$$n_{e(h)}=g\int\frac{d^2k}{(2\pi)^2}\frac{\hbar v_{\boldsymbol{k},e(h)}k}{2}\left(-\frac{\partial f^{(0)}_{\mathbf{k},e(h)}}{\partial E_{\mathbf{k},e(h)}}\right),\tag{B5}$$

as electron(hole) effective mass density, stress-energy tensor and carrier density, respectively.

In terms of these quantities, the Navier-Stokes equation becomes

$$\partial_t\left(\rho_{m,e(h)}\boldsymbol{u}_{e(h)}\right)+\nabla_\alpha\Pi^{\alpha\beta}_{e(h)}+\lambda en_{e(h)}\boldsymbol{E}=-n_{e(h)}\frac{\boldsymbol{u}_{e(h)}-\boldsymbol{u}_{h(e)}}{\mu_{C,e(h)}}-n_{e(h)}\frac{\boldsymbol{u}_{e(h)}}{\mu_{\mathrm{dis},e(h)}}\tag{B6}$$

where $\mu_{C,e(h)}$ is electron(hole) Coulomb-interaction mobility and $\mu_{\mathrm{dis},e(h)}$ is electron(hole) disorder mobility.

Adding electron and hole component (and assume $\boldsymbol{u}_e\approx\boldsymbol{u}_h\equiv\boldsymbol{u}$ which is valid in hydrodynamic regime) and using conservation of momentum

$$\partial_t\left(\rho_m\boldsymbol{u}\right)+\nabla_\alpha\Pi^{\alpha\beta}+en\boldsymbol{E}=-\left(\frac{n_e}{\mu_{\mathrm{dis},e}}+\frac{n_h}{\mu_{\mathrm{dis},h}}\right)\boldsymbol{u}\tag{B7}$$

where we have defined $\rho_m=\rho_{m,e}+\rho_{m,h},\Pi^{\alpha\beta}=\Pi^{\alpha\beta}_e+\Pi^{\alpha\beta}_h$ and $n=n_e-n_h$.

In terms of disorder resistivity $\rho_{\mathrm{dis},e(h)}$

$$\partial_t\left(\rho_m\boldsymbol{u}\right)+\nabla_\alpha\Pi^{\alpha\beta}+en\boldsymbol{E}=-e\left(\rho_{\mathrm{dis},e}n_e^2+\rho_{\mathrm{dis},h}n_h^2\right)\boldsymbol{u}\tag{B8}$$

The first term can be written as

$$\partial_t\left(\rho_m\boldsymbol{u}\right)=\rho_m\partial_t\boldsymbol{u}+\boldsymbol{u}\partial_t\rho_m,\tag{B9}$$

Meanwhile, the second term can be written as

$$\Pi^{\alpha\beta}=P\delta^{\alpha\beta}+\rho_m u^\alpha u^\beta+\delta\Pi^{\alpha\beta}\tag{B10}$$

where $P$ is pressure and $\delta\Pi^{\alpha\beta}$ is the dissipative correction to the stress-energy tensor.

Here we show how to derive this relation for bilayer graphene (see Ref. [65] for the case of monolayer). We start from the definition of electron(hole) stress-energy tensor

$$\begin{aligned}\Pi^{\alpha\beta}_{e(h),\mathrm{BLG}}&=g\int\frac{d^2k}{(2\pi)^2}\hbar v^\alpha_{\boldsymbol{k},e(h)}k^\beta f_{\mathbf{k},e(h)}\\&=m^\star n_{e(h)}\left\langle v^\alpha_{\boldsymbol{k},e(h)}v^\beta_{\boldsymbol{k},e(h)}\right\rangle,\end{aligned}\tag{B11}$$

where we have used notation $\langle\ldots\rangle=g\int\frac{d^2k}{(2\pi)^2}\left(\ldots\right)f_{\mathbf{k},e(h)}/g\int\frac{d^2k}{(2\pi)^2}f^{(0)}_{\mathbf{k},e(h)}$.

We can write $\left\langle v^{\alpha}_{\boldsymbol{k},e(h)} v^{\beta}_{\boldsymbol{k},e(h)} \right\rangle$ as

$$\begin{aligned}
\left\langle v^{\alpha}_{\boldsymbol{k},e(h)} v^{\beta}_{\boldsymbol{k},e(h)} \right\rangle &= \left\langle \left( v^{\alpha}_{\boldsymbol{k},e(h)} - u^{\alpha}_{e(h)} \right) \left( v^{\beta}_{\boldsymbol{k},e(h)} - u^{\beta}_{e(h)} \right) \right\rangle + \left\langle v^{\alpha}_{\boldsymbol{k},e(h)} u^{\beta}_{e(h)} \right\rangle + \left\langle u^{\alpha}_{e(h)} v^{\beta}_{\boldsymbol{k},e(h)} \right\rangle - \left\langle u^{\alpha}_{e(h)} u^{\beta}_{e(h)} \right\rangle \\
&= \left\langle \left( v^{\alpha}_{\boldsymbol{k},e(h)} - u^{\alpha}_{e(h)} \right) \left( v^{\beta}_{\boldsymbol{k},e(h)} - u^{\beta}_{e(h)} \right) \right\rangle + \left\langle v^{\alpha}_{\boldsymbol{k},e(h)} \right\rangle u^{\beta}_{e(h)} + u^{\alpha}_{e(h)} \left\langle v^{\beta}_{\boldsymbol{k},e(h)} \right\rangle - u^{\alpha}_{e(h)} u^{\beta}_{e(h)} \\
&= \left\langle \left( v^{\alpha}_{\boldsymbol{k},e(h)} - u^{\alpha}_{e(h)} \right) \left( v^{\beta}_{\boldsymbol{k},e(h)} - u^{\beta}_{e(h)} \right) \right\rangle + u^{\alpha}_{e(h)} u^{\beta}_{e(h)} \qquad \text{(B12)}
\end{aligned}$$

where we have used the relation between drift velocity $\boldsymbol{u}_{e(h)}$ and group velocity $\boldsymbol{v}_{\boldsymbol{k},e(h)}$, i. e. $\boldsymbol{u}_{e(h)} = \langle \boldsymbol{v}_{\boldsymbol{k},e(h)} \rangle$.

Hence

$$\Pi^{\alpha\beta}_{e(h),\mathrm{BLG}} = m^{\star} n_{e(h)} \left\langle \left( v^{\alpha}_{\boldsymbol{k},e(h)} - u^{\alpha}_{e(h)} \right) \left( v^{\beta}_{\boldsymbol{k},e(h)} - u^{\beta}_{e(h)} \right) \right\rangle + m^{\star} n_{e(h)} u^{\alpha}_{e(h)} u^{\beta}_{e(h)} \qquad \text{(B13)}$$

The first term can be decompose as the sum of diagonal component due to equilibrium distribution function and off-diagonal component due to deviation from equilibrium, i. e.

$$\Pi^{\alpha\beta}_{e(h),\mathrm{BLG}} = P_{e(h)} \delta_{\alpha\beta} + \delta\Pi^{\alpha\beta}_{e(h)} + m^{\star} n_{e(h)} u^{\alpha}_{e(h)} u^{\beta}_{e(h)} \qquad \text{(B14)}$$

Hence the spatial derivative of the total stress energy tensor in hydrodynamic regime becomes

$$\nabla_{\alpha} \Pi^{\alpha\beta} = \nabla P + \rho_m \left( \boldsymbol{u} \cdot \nabla \right) \boldsymbol{u} + \boldsymbol{u} \nabla \cdot \left( \rho_m \boldsymbol{u} \right) + \nabla_{\alpha} \left( \delta\Pi^{\alpha\beta} \right), \qquad \text{(B15)}$$

leading to

$$\rho_m \left[ \partial_t + \left( \boldsymbol{u} \cdot \nabla_{\boldsymbol{r}} \right) \right] \boldsymbol{u} = -en\boldsymbol{E} - \nabla_{\alpha} \left( \delta\Pi^{\alpha\beta} \right) - \nabla_{\boldsymbol{r}} P - \left[ \partial_t \rho_m + \nabla_{\boldsymbol{r}} \cdot \left( \rho_m \boldsymbol{u} \right) \right] \boldsymbol{u} - e \left( \rho_{\mathrm{dis},e} n_e^2 + \rho_{\mathrm{dis},h} n_h^2 \right) \boldsymbol{u} \qquad \text{(B16)}$$

The dissipative correction to the Navier-Stokes equation governed by rotational invariance and in 2D in the absence of external magnetic field is given by

$$\delta\Pi^{\alpha\beta} = -\eta \left( \nabla^{\alpha} u^{\beta} + \nabla^{\beta} u^{\alpha} - \delta^{\alpha\beta} \nabla \cdot \boldsymbol{u} \right) - \zeta \delta^{\alpha\beta} \nabla \cdot \boldsymbol{u}, \qquad \text{(B17)}$$

where $\eta$ and $\zeta$ are the shear and bulk viscosity, respectively.

Substituting the dissipative correction, we get

$$\rho_m \left[ \partial_t + \left( \boldsymbol{u} \cdot \nabla \right) \right] \boldsymbol{u} = -en\boldsymbol{E} + \eta \nabla^2 \boldsymbol{u} + \left( \frac{\eta}{2} + \zeta \right) \nabla \left( \nabla \cdot \boldsymbol{u} \right) - \nabla P - \left[ \partial_t \rho_m + \nabla \cdot \left( \rho_m \boldsymbol{u} \right) \right] \boldsymbol{u} - e \left( \rho_{\mathrm{dis},e} n_e^2 + \rho_{\mathrm{dis},h} n_h^2 \right) \boldsymbol{u}, \qquad \text{(B18)}$$

which in the absence of bulk viscosity becomes eq. 4 in the main text.

---